\documentclass{article}
\usepackage[utf8]{inputenc}
\usepackage{spconf,amsmath,amssymb,graphicx,booktabs,hyperref,afterpage}
\hypersetup{hidelinks}
\title{LIKELIHOOD-CONSTRAINED ACOUSTIC RERANKING FOR\\TRAINING-FREE HALLUCINATION MITIGATION IN LLM-BASED ASR}
\name{Jiasheng Kuang$^{\star}$, Linru Zheng, Hongjin Song, Zhaoqi Cui, Song Li$^{\dagger}$}
\address{Shengwang}

\begin{document}
\ninept
\maketitle
\begin{abstract}
Large language model (LLM)-based automatic speech recognition (ASR) systems achieve strong performance on conventional speech data by leveraging powerful linguistic priors and multilingual capabilities. However, under challenging conditions, these priors can override acoustic evidence, resulting in unintended translation, instruction execution, repetition, or catastrophic deletion. We propose Likelihood-Constrained Acoustic Reranking (LCAR), a training-free decoding method that improves acoustic grounding while preserving support from the base model. At each decoding step, LCAR first retains tokens whose base-model likelihood falls within a margin of the greedy token, then reranks them using an acoustic compatibility score computed from attention-pooled audio embeddings and the existing LM head. By restricting acoustic intervention to plausible, model-supported alternatives, LCAR requires no additional training, external detector, reference transcript, or auxiliary model at inference. We evaluate LCAR on four LLM-based ASR systems using human-audited TTS and open-source speech challenge suites. At $\delta=0.60$, LCAR removes 38.8--57.1\% of detector-identified hallucination failures while largely maintaining WER/CER on standard open-source test sets.
\end{abstract}

\section{Introduction}
Modern ASR systems increasingly combine audio encoders with autoregressive language decoders \cite{qwen2,qwen3,kimi,glm}. Their linguistic priors improve fluency and multilingual coverage, but can override the transcription objective. A code-switched phrase, here limited to Chinese--English switching, may be translated into one language; a conflicting prompt may override the ASR instruction; or an audible ``please translate'' may be obeyed instead of transcribed.

Increasing acoustic influence can redirect a hallucinated trajectory, but unconstrained reranking can also disrupt recognition and EOS decisions. LCAR admits only tokens within $\delta$ nats of the greedy token and uses an internal acoustic score to choose among them. It exactly matches greedy decoding at $\delta=0$ and requires no detector, reference, auxiliary model, or additional training at inference.

Prior studies characterize ASR hallucinations under acoustic perturbations, non-speech, clinical
audio, and distribution shift \cite{frieske,whispercd,careless,lost}. Speech Hallucination Overview (SHALLOW) organizes recurring failure patterns, while HALAS provides span-level annotations on real speech \cite{shallow,halas}. Our suites instead freeze model-specific code-switching and instruction-induced failures to support paired decoding experiments.

Reference-free detection has used text statistics, decoder states, and cross-modal attention [12, 13]. Mitigation methods include phrase suppression, attention-head adaptation, policy optimization, and contrastive decoding [7, 14, 15, 16, 17, 18]. Whisper-CD contrasts clean logits with noise, silence, and temporal-shift negatives. LCAR uses the unperturbed forward pass and restricts attention-based reranking with the
model likelihood; no inference-time detector is needed.

Our contributions are threefold: (1) a grounding-based account of translation/transliteration, instruction execution, unsupported repetition, and catastrophic deletion in LLM-based ASR; (2) LCAR, a training-free decoder that bounds candidate likelihood before acoustic reranking; and (3) two human-audited, 500-utterance challenge suites, one TTS-based and one based on open-source speech, both of which are publicly available on Hugging Face.

\section{Benchmark Construction}
\subsection{Hallucination Categories}
We define an acoustic-grounding hallucination as a catastrophic decoding trajectory in which linguistic or task priors override available acoustic evidence. Translation or transliteration converts a meaning-bearing span; instruction execution obeys an audible or textual command; repetition adds a sustained loop unsupported by the audio; and catastrophic deletion removes most spoken content. Ordinary omissions remain ASR errors. CS-HAL targets mixed-language transformation, while Prompt-HAL and Spoken-HAL target textual and audible instructions, respectively. Prompt-HAL pairs an attacked prompt with a normal prompt for the same audio. Because GLM has no documented instruction channel, Prompt-HAL is not evaluated for this model.

\subsection{Challenge Suites}
IndexTTS2 generates coherent and well-balanced Mandarin-English utterances. OpenSpeech integrates original code-switched audio from TALCS, ASCEND, NTUML2021, CS-Dialogue, and YODAS with corresponding text-based attacks on the same audio and synthesized spoken instructions.
Each released suite contains 500 human-audited source utterances and is available at \url{https://huggingface.co/aguangguang}. Candidate pools are decoded and filtered separately for each model. We then fix 400 positive examples for every available model--suite--category track, yielding 8800 frozen records across 22 tracks. These stress-test sets are designed to probe failure modes rather than estimate their real-world prevalence.

\begin{figure*}[t]
\centering
\includegraphics[width=0.98\textwidth]{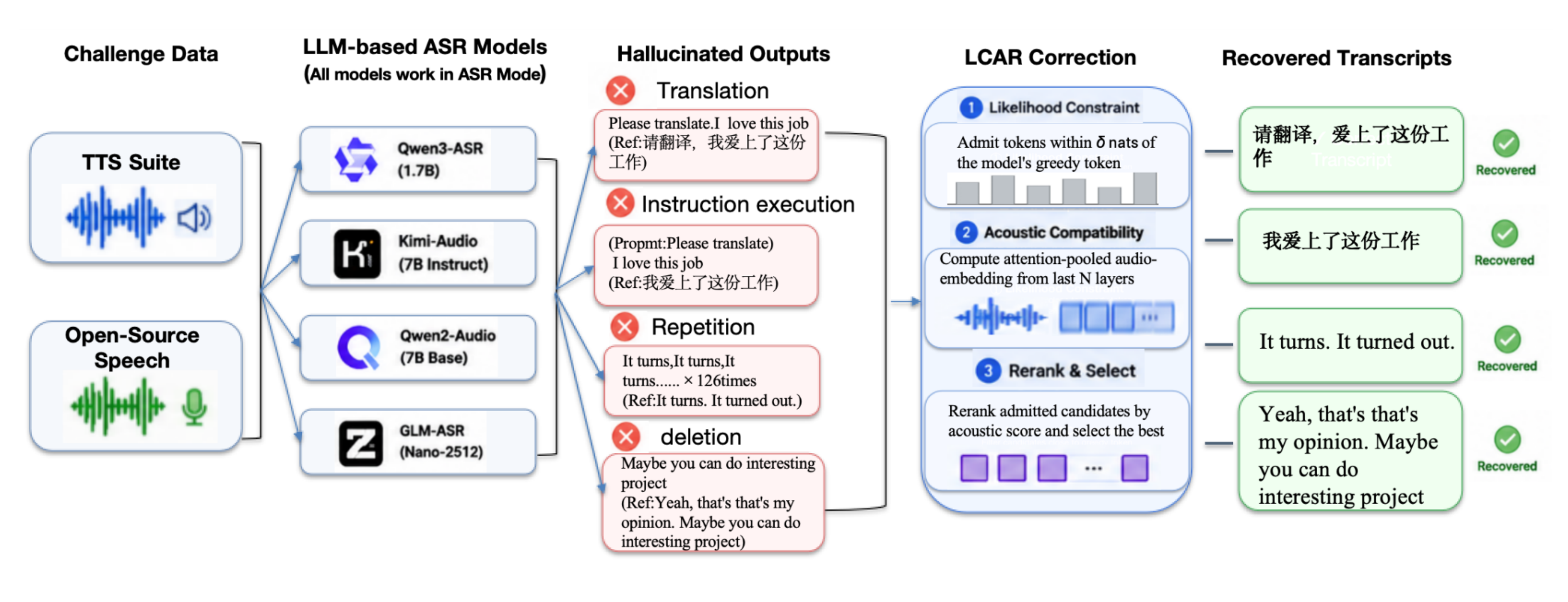}
\caption{Four acoustic-grounding failure modes and the LCAR decoding pipeline. LCAR reranks near-greedy tokens using attention-derived acoustic compatibility.}
\label{fig:pipeline}
\end{figure*}

\afterpage{%
\begin{table*}[!t]
\caption{Reference-audited outcomes at $\delta=0.60$. Values are counts with percentages of detected events in parentheses; TTS denotes the dataset generated with IndexTTS. Faithful recovery has CUER $\le10\%$; Section 4.1 defines the mixed-language content units and audit procedure.}
\label{tab:outcomes}
\centering\small
\renewcommand{\arraystretch}{0.80}
\begin{tabular*}{\textwidth}{@{\extracolsep{\fill}}llrrrrr}
\toprule Model & Suite & \shortstack{Detected\\events} & \multicolumn{4}{c}{Outcome after detector exit} \\
\cmidrule(lr){4-7}
& & & \multicolumn{1}{c}{\shortstack{Faithful\\recovery $\uparrow$}} & \multicolumn{1}{c}{\shortstack{Ordinary\\ASR error}} & \multicolumn{1}{c}{\shortstack{Residual\\HIR $\downarrow$}} & \multicolumn{1}{c}{\shortstack{New-type\\HIR $\downarrow$}} \\
\midrule
Qwen3 & TTS & 526 & 306 (58.2\%) & 209 (39.7\%) & 9 (1.7\%) & 2 (0.4\%) \\
& OpenSpeech & 651 & 236 (36.3\%) & 271 (41.6\%) & 111 (17.1\%) & 33 (5.1\%) \\
\midrule
Qwen2 Base & TTS & 134 & 55 (41.0\%) & 70 (52.2\%) & 9 (6.7\%) & 0 (0.0\%) \\
& OpenSpeech & 797 & 67 (8.4\%) & 452 (56.7\%) & 151 (18.9\%) & 127 (15.9\%) \\
\midrule
Kimi & TTS & 341 & 92 (27.0\%) & 218 (63.9\%) & 20 (5.9\%) & 11 (3.2\%) \\
& OpenSpeech & 635 & 60 (9.4\%) & 429 (67.6\%) & 85 (13.4\%) & 61 (9.6\%) \\
\midrule
GLM & TTS & 346 & 241 (69.7\%) & 94 (27.2\%) & 10 (2.9\%) & 1 (0.3\%) \\
& OpenSpeech & 567 & 95 (16.8\%) & 337 (59.4\%) & 78 (13.8\%) & 57 (10.1\%) \\
\bottomrule
\end{tabular*}
\par\raggedright\small Percentages in parentheses are conditional on detected events. Residual HIR retains the initial failure family; new-type HIR changes family. Outputs without an identifiable final family (49 cases) are conservatively counted as residual.
\end{table*}
\begin{figure*}[!t]
\centering
\includegraphics[width=0.86\textwidth]{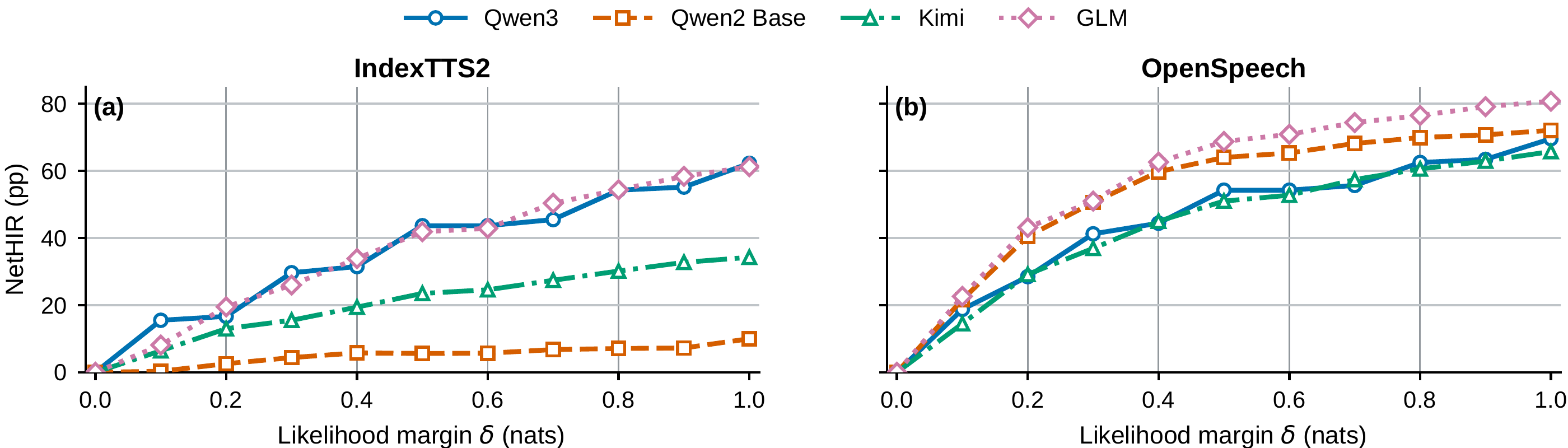}
\caption{NetHIR under Last-4 LCAR on (a) IndexTTS2 and (b) OpenSpeech. NetHIR subtracts induction on equally sized negative controls from detector exits on frozen positive tracks.}
\label{fig:nethir}
\end{figure*}
\begin{figure*}[!t]
\centering
\includegraphics[width=0.86\textwidth]{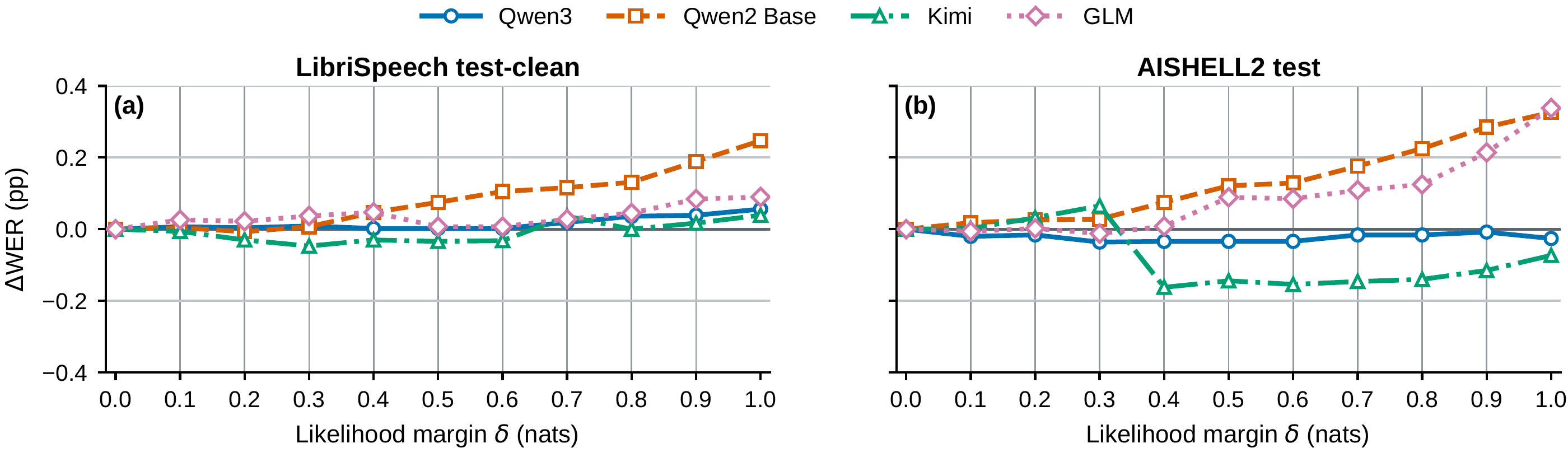}
\caption{Normal-speech side effects across the margin sweep. Curves show paired changes from matched-greedy decoding on LibriSpeech test-clean and AISHELL2 test; negative values indicate lower error.}
\label{fig:sideeffects}
\end{figure*}
}

\subsection{Semantic Screening}
Using frozen structured prompts, Qwen3-32B [25] performs a two-stage adjudication process. In the first stage, it aligns the reference and hypothesis and assigns a provisional failure category. In the second stage, it verifies the supporting evidence, checks whether the category-specific criterion is satisfied, and assesses whether the observed failure is attributable to the Prompt-HAL attack.
Translation and transliteration are defined as cases in which a meaning-bearing span crosses a language boundary or a script boundary, respectively. Instruction execution occurs when the model follows an audible or textual attack rather than transcribing it.
Repetition must be sustained and unsupported by the acoustic signal, whereas severe deletion must either omit most of the spoken content or collapse to an instruction prefix.

Formatting differences, conventional name or acronym renderings, and local sound-alike substitutions are excluded. A detected event is counted only if the output differs from the same-run greedy-decoding output after text normalization.

One annotator independently reviewed a fixed-seed random sample of 400 detector positives (200 per suite). Sampling covered every available model--suite--category track and was stratified by detector label. Precision is 93.3\% (373/400); the 27 rejected cases are ordinary ASR errors. Stratum-weighted precision is 93.9\% (95\% CI: 93.1--94.7\%).

\section{Method}
LCAR limits token admission by base-model likelihood, then ranks the admitted tokens by audio-conditioned compatibility. We distinguish two uses of the LM head: (1) the standard LM head, which produces the base model's next-token distribution from the decoder states; and (2) the same LM head, reused as a projector for the pooled audio embedding, which yields an acoustic compatibility score over the vocabulary. While both operations share the same output projection weights, the former operates on the language-model state, whereas the latter operates on an attention-pooled acoustic context.
\subsection{Acoustic Compatibility}
At decoding step $t$, the base-LM logit vector is $\ell_t^m\in\mathbb{R}^{|V|}$, where superscript $m$ denotes the base language-model stream, $V$ is the vocabulary, and $|V|$ is its size. For a vocabulary token $y$, $q_t^m(y)$ is its log probability; $\operatorname{softmax}(\ell_t^m)_y$ selects its $y$-th probability; and $y_t^g$ is the greedy token chosen by $\arg\max$:
\begin{equation}
 q_t^m(y)=\log\operatorname{softmax}(\ell_t^m)_y,\qquad y_t^g=\arg\max_y q_t^m(y).
\end{equation}
The audio encoder and projector provide $N_a$ audio embeddings $e^a_{1:N_a}$ to the LM, where superscript $a$ denotes acoustic quantities and index $i\in\{1,\ldots,N_a\}$ identifies an audio position. Let $L$ be the last decoder-layer index, $n$ the retained number of final layers, $L_n=\{L-n+1,\ldots,L\}$ their index set, and $H$ the number of attention heads. $A^{(\ell,h)}_{t,i}$ is the attention from generated position $t$ to input position $i$ in layer $\ell\in L_n$ and head $h$; averaging these $nH$ values gives
\begin{equation}
 w_{t,i}=\frac{1}{nH}\sum_{\ell\in L_n}\sum_{h=1}^H A^{(\ell,h)}_{t,i}.
\end{equation}
We renormalize the attention weights over the audio positions, $\tilde w_{t,i}=w_{t,i}/\sum_{j=1}^{N_a}w_{t,j}$, where $j$ ranges over those positions. The pooled vector $c_t$ is the audio context at step $t$, and $\operatorname{LN}$ is the LM output layer normalization:
\begin{equation}
 c_t=\operatorname{LN}\left(\sum_{i=1}^{N_a}\tilde w_{t,i}e^a_i\right),\qquad q_t^a(y)=\log\operatorname{softmax}(Wc_t)_y.
\end{equation}
Here $W$ is the LM-head matrix: the existing output projection from the normalized representation to $|V|$ vocabulary logits. Thus $Wc_t$ scores every token from the pooled acoustic context, and $q_t^a(y)$ is an acoustic compatibility score rather than a calibrated token probability.

\subsection{Likelihood-Constrained Decoding}
Applying $q_t^a$ to the full vocabulary can select a token with negligible LM support. LCAR first forms the candidate set $C_t(\delta)$, where $\delta\ge0$ is the likelihood margin in nats:
\begin{equation}
 C_t(\delta)=\{y:q_t^m(y)\ge q_t^m(y_t^g)-\delta\}.
\end{equation}
It then chooses the acoustically most compatible admitted token:
\begin{equation}
 y_t^{\mathrm{LCAR}}=\arg\max_{y\in C_t(\delta)}q_t^a(y).
\end{equation}
Thus every selected token is within $\delta$ nats in log probability of the greedy token. At $\delta=0$, $C_t$ contains only $y_t^g$, so LCAR equals greedy decoding. EOS is never overridden; deletion is repaired only through an earlier prefix change. The method reuses encoder states and adds averaging over $n$ layers plus one LM-head projection per step.

\section{Experimental Setup}
\subsection{Models and Evaluation}
{\sloppy
We evaluate Qwen3-ASR-1.7B, Qwen2-Audio-7B Base, Kimi-Audio-7B-Instruct, and GLM-ASR-Nano-2512 with official prompts and preprocessing. The main sweep uses $n=4$ and eager attention extraction. Every track contains 400 matched-greedy positives and 400 negative controls. Normal-speech evaluation uses 2,620 LibriSpeech test-clean utterances and 5,000 AISHELL2 utterances, scored by WER and CER.
\par}

Mixed-language content units comprise Han characters, case-folded English words, and digit strings. For reference units $R$, substitutions $S$, deletions $D$, and insertions $I$,
\begin{equation}
 \mathrm{CUER}=\frac{S+D+I}{|R|}.
\end{equation}
Let $h_i(\delta)\in\{0,1\}$ be the detector decision. Paired transition rates are
\begin{equation}
 \mathrm{HIR\mathchar`-Fix}(\delta)=\frac{\sum_i\mathbf{1}[h_i(0)=1\land h_i(\delta)=0]}{\sum_i\mathbf{1}[h_i(0)=1]},
\end{equation}
\begin{equation}
 \begin{aligned}
 \mathrm{HIR\mathchar`-Induce}(\delta)&=\frac{\sum_i\mathbf{1}[h_i(0)=0\land h_i(\delta)=1]}{\sum_i\mathbf{1}[h_i(0)=0]},\\
 \mathrm{NetHIR}&=\mathrm{HIR\mathchar`-Fix}-\mathrm{HIR\mathchar`-Induce}.
 \end{aligned}
\end{equation}
HIR-Fix and HIR-Induce are the positive-to-negative and
negative-to-positive detector transition rates. A Qwen3 pilot selected $\delta=0.60$ before benchmark construction; this
threshold retains tokens with at least $e^{-0.60}=54.9\%$ of the
greedy token probability. The full grid tests sensitivity to a
shared margin. HIR-Induce 95\% CIs use 10,000 paired bootstrap resamples, stratified by track and clustered by source
utterance. Candidate-set size and token-switch rate describe
the amount of intervention.

\begin{figure}[t]
\centering
\includegraphics[width=0.80\columnwidth]{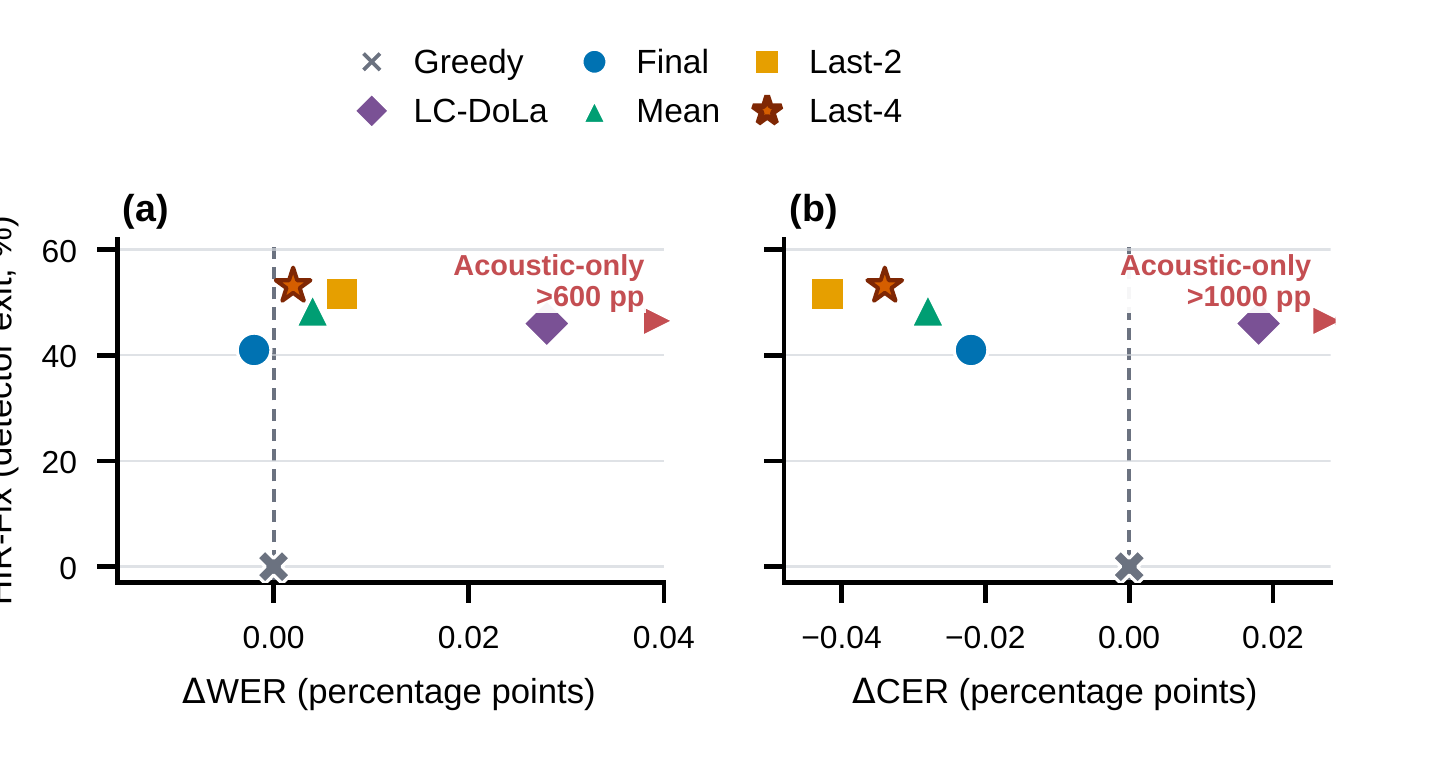}
\caption{Qwen3 trade-offs at $\delta=0.60$: HIR-Fix versus paired (a) $\Delta$WER and (b) $\Delta$CER. Acoustic-only diverges beyond 600 percentage points because insertion errors are unbounded.}
\label{fig:tradeoff}
\makeatletter
\def\@captype{table}
\makeatother
\caption{Qwen3 ablation at $\delta=0.60$ on 800 positives (400\protect\\ per suite). HIR-Fix: TTS/OpenSpeech/all; Switch: token-\protect\\ switch rate. Arrows show preference.}
\label{tab:ablation}
{\small
\setlength{\tabcolsep}{1pt}
\renewcommand{\arraystretch}{0.80}
\begin{tabular*}{\columnwidth}{@{\extracolsep{\fill}}lrrr}
\toprule Variant & \shortstack{HIR-Fix\\TTS/OpenSpeech/all $\uparrow$} & \shortstack{$\Delta$WER/\\$\Delta$CER $\downarrow$} & \shortstack{$|C_t|$/\\Switch $\downarrow$} \\
\midrule
\multicolumn{4}{l}{Baselines} \\
Greedy & 0/0/0 & +0.000/+0.000 & 1.000/0.0 \\
LC-DoLa & 35.3/56.8/46.0 & +0.028/+0.018 & 1.071/3.3 \\
\midrule
\multicolumn{4}{l}{LCAR variants} \\
Final & 35.8/46.3/41.0 & -0.002/-0.022 & 1.073/3.3 \\
Last-2 & 48.8/54.5/51.6 & +0.007/-0.042 & 1.074/3.9 \\
\textbf{Last-4} & \textbf{49.5/56.8/53.1} & \textbf{+0.002/-0.034} & \textbf{1.076/4.0} \\
Mean pooling & 42.8/54.3/48.5 & +0.004/-0.028 & 1.075/3.9 \\
\midrule
\multicolumn{4}{l}{Constraint ablation} \\
Acoustic-only & 17.0/76.0/46.5 & \shortstack{$>+600/$\\$>+1000$} & full/34.1 \\
\bottomrule
\end{tabular*}
\par\small $\delta=0$ is zero by construction.
}
\end{figure}

\section{Results and Discussion}
\subsection{Main Results}
NetHIR is positive for every model--suite pair (Fig.~2). At $\delta=0.60$, it ranges from 5.7 to 43.7 points on IndexTTS2 and from 52.8 to 70.9 points on OpenSpeech; the overall range at $\delta=1.0$ is 10.0--80.8 points. As the margin grows, more near-greedy alternatives enter the candidate set and NetHIR generally rises. Qwen2 Base gains less on IndexTTS2, showing a clear dependence on both the model and the failure set.

\subsection{Normal-Speech Evaluation}
At $\delta=0.60$, the WER/CER changes are +0.002\%/-0.034\% for Qwen3, -0.032\%/-0.154\% for Kimi, +0.105\%/+0.129\% for Qwen2 Base, and +0.007\%/+0.085\% for GLM (Fig.~3). Side effects remain small at conservative margins but widen for Qwen2 Base and GLM. Larger $\delta$ admits progressively less likely tokens, which is consistent with both improved trajectory recovery and architecture-dependent recognition costs.

\subsection{Ablation and Repair Analysis}
The Qwen3 ablation examines late-layer aggregation, temporal pooling, and the likelihood constraint on 800 positives (400 per suite). LC-DoLa shares $C_t(0.60)$ and selects among 1/4-, 1/2-, and 3/4-depth layers by maximum Jensen--Shannon divergence [17]; Final, Last-2, and Last-4 use $n=1,2,4$.
Last-4 attains 53.1\% HIR-Fix versus 51.6\% for Last-2, 48.5\% for mean pooling, and 41.0\% for the final layer, with +0.002/-0.034 normal-speech changes (Table 2; Fig.~4). Its 1.076 mean candidate-set size and 4.0\% switch rate indicate sparse corrections.
The likelihood constraint is essential: LC-DoLa reaches 46.0\% HIR-Fix, whereas acoustic-only selection causes more than 600/1,000 percentage-point WER/CER increases. Of the 3,997 detected events, 28.8\% yield faithful recovery and 52.0\% ordinary ASR errors; a detector exit is therefore not synonymous with exact recovery.
Table 1 shows a suite gap: TTS yields 51.5\% faithful recovery and 4.6\% residual/new-type HIR, versus 17.3\% and 26.5\% on OpenSpeech. LCAR redirects many hallucinated trajectories into ordinary ASR outputs, so we report both reference accuracy and NetHIR.

\section{Conclusion}
In this work, we identify several acoustic-grounding hallucination patterns in LLM-based ASR, including unintended translation or transliteration, instruction execution, unsupported repetition, and catastrophic deletion. To mitigate these failures, we propose Likelihood-Constrained Acoustic Reranking (LCAR), a training-free decoding method that first restricts candidates to tokens supported by the base-model likelihood and then selects among them using attention-derived acoustic compatibility. Across four LLM-based ASR systems and two human-audited challenge suites, LCAR effectively reduces detector-identified hallucinations while largely preserving recognition performance on standard open-source test sets. Future work will extend LCAR to efficient attention backends, adaptive likelihood margins, streaming ASR, and broader linguistic and acoustic conditions.

\end{document}